\documentclass[preprint]{ptephy_v1}

\preprintnumber{XXXX-XXXX} 
\usepackage{hyperref}

\usepackage{multirow}
\usepackage{ragged2e}

\begin{document}

\title{Pulse Shape Discrimination in JSNS$^2$}


\author[1,2]{T. Dodo}
\author[3]{M. K. Cheoun}
\author[4]{J. H. Choi}
\author[5]{J. Y. Choi}
\author[6]{J. Goh}
\author[7]{K. Haga}
\author[7]{M. Harada}
\author[2,7]{S. Hasegawa}
\author[6]{W. Hwang}
\author[8]{T. Iida}
\author[5]{H. I. Jang}
\author[9]{J. S. Jang}
\author[10]{K. K. Joo}
\author[11]{D. E. Jung}
\author[12]{S. K. Kang}
\author[7]{Y. Kasugai}
\author[13]{T. Kawasaki}
\author[14]{E. J. Kim}
\author[10]{J. Y. Kim}
\author[15]{S. B. Kim}
\author[16]{W. Kim}
\author[7]{H. Kinoshita}
\author[13]{T. Konno}
\author[17]{D. H. Lee}
\author[10]{I. T. Lim}
\author[18]{C. Little}
\author[18]{E. Marzec}
\author[17]{T. Maruyama}
\author[7]{S. Masuda}
\author[7]{S. Meigo}
\author[10]{D. H. Moon}
\author[19]{T. Nakano}
\author[20]{M. Niiyama}
\author[17]{K. Nishikawa}
\author[4]{M. Y. Pac}
\author[10]{H. W. Park}
\author[16]{J. S. Park}
\author[10]{R. G. Park}
\author[21]{S. J. M. Peeters}
\author[22]{C. Rott}
\author[7]{K. Sakai}
\author[7]{S. Sakamoto}
\author[19]{T. Shima}
\author[17]{C. D. Shin \thanks{Corresponding author}}
\author[18]{J. Spitz}
\author[1]{F. Suekane}
\author[19]{Y.Sugaya}
\author[7]{K. Suzuya}
\author[8]{Y.Takeuchi}
\author[7]{Y.Yamaguchi}
\author[23]{M. Yeh}
\author[4]{I. S. Yeo}
\author[6]{C. Yoo}
\author[11]{I. Yu}

\affil[1]{Research Center for Neutrino Science, Tohoku University, 6-3 Azaaoba, Aramaki, Aoba-ku, Sendai 980-8578, Japan}
\affil[2]{Advanced Science Research Center, JAEA, 2-4 Shirakata, Tokai-mura, Naka-gun, Ibaraki 319-1195, Japan}
\affil[3]{Department of Physics and OMEG Institute, Soongsil University, 369 Sangdo-ro, Dongjak-gu, Seoul, 06978, Korea}
\affil[4]{Laboratory for High Energy Physics, Dongshin University, 67, Dongshindae-gil, Naju-si, Jeollanam-do, 58245, Korea}
\affil[5]{Department of Fire Safety, Seoyeong University, 1 Seogang-ro, Buk-gu, Gwangju, 61268, Korea}
\affil[6]{Department of Physics, Kyung Hee University, 26, Kyungheedae-ro, Dongdaemun-gu, Seoul 02447, Korea}
\affil[7]{J-PARC Center, JAEA, 2-4 Shirakata, Tokai-mura, Naka-gun, Ibaraki 319-1195, Japan}
\affil[8]{Faculty of Pure and Applied Sciences, University of Tsukuba,\\ Tennodai 1-1-1, Tsukuba, Ibaraki, 305-8571, Japan}
\affil[9]{Department of Physics and Photon Science, Gwangju Institute of Science and Technology, 123 Cheomdangwagi-ro, Buk-gu, Gwangju, 61005, Korea}
\affil[10]{Department of Physics, Chonnam National University, 77, Yongbong-ro, Buk-gu, Gwangju, 61186, Korea}
\affil[11]{Department of Physics, Sungkyunkwan University, 2066, Seobu-ro, Jangan-gu, Suwon-si, Gyeonggi-do, 16419, Korea}
\affil[12]{School of Liberal Arts, Seoul National University of Science and Technology, 232 Gongneung-ro, Nowon-gu, Seoul, 139-743, Korea}
\affil[13]{Department of Physics, Kitasato University, 1 Chome-15-1 Kitazato, Minami Ward, Sagamihara, Kanagawa, 252-0329, Japan}
\affil[14]{Division of Science Education, Jeonbuk National University, 567 Baekje-daero, Deokjin-gu, Jeonju-si, Jeollabuk-do, 54896, Korea}
\affil[15]{School of Physics, Sun Yat-sen (Zhongshan) University, Haizhu District, Guangzhou, 510275, China}
\affil[16]{Department of Physics, Kyungpook National University, 80 Daehak-ro, Buk-gu, Daegu, 41566, Korea}
\affil[17]{High Energy Accelerator Research Organization (KEK), 1-1 Oho, Tsukuba, Ibaraki, 305-0801, Japan}
\affil[18]{University of Michigan, 500 S. State Street, Ann Arbor, MI 48109, U.S.A.}
\affil[19]{Research Center for Nuclear Physics, Osaka University, 10-1 Mihogaoka, Ibaraki, Osaka, 567-0047, Japan}
\affil[20]{Department of Physics, Kyoto Sangyo University, Motoyama, Kamigamo, Kita-Ku, Kyoto-City, 603-8555, Japan}
\affil[21]{Department of Physics and Astronomy, University of Sussex, Falmer, Brighton, BN1 9RH, U.K.}
\affil[22]{Department of Physics and Astronomy, University of Utah, 201 Presidents' Cir, Salt Lake City, UT 84112, U.S.A}
\affil[23]{Brookhaven National Laboratory, Upton, NY 11973-5000, U.S.A.}


\begin{abstract}%
 JSNS$^2$ (J-PARC Sterile Neutrino Search at J-PARC Spallation Neutron Source)
  is an experiment that is searching for sterile neutrinos via the observation
  of $\bar{\nu}_{\mu} \rightarrow \bar{\nu}_e$ appearance oscillations
  using neutrinos with muon decay-at-rest. For this search, rejecting cosmic-ray-induced neutron events by Pulse Shape Discrimination (PSD) is essential 
  because the JSNS$^2$ detector is located above ground, on the third floor of the building.
  We have achieved 94.95$\pm$0.15\% rejection of neutron events while keeping 92.82$\pm$1.77\% of signal, electron-like events using a data driven likelihood method. This article will report the PSD technique using the full fiducial volume of the JSNS$^2$ detector.

\end{abstract}

\subjectindex{C30, H14, H16}

\maketitle

\section{\label{sec:intro} Introduction }

The existence of sterile neutrinos has been an important issue in the
field of neutrino physics for over 20 years.
The experimental results from \cite{CITE:LSND, CITE:BEST, CITE:MiniBooNE2018, CITE:REACTOR} could be interpreted as indications of the
existence of sterile neutrinos with mass-squared differences of
around 1~eV$^2$.

The JSNS$^2$ experiment, proposed in 2013~\cite{CITE:JSNS2proposal}, 
searches for short-baseline neutrino oscillations at the Material and Life science experimental Facility (MLF)
in J-PARC.
The facility provides an intense and high-quality neutrino source
with $1.8\times10^{14}$ $\nu$/year/cm$^2$ from
muon decay-at-rest ($\mu$DAR). 
These neutrinos are produced by impinging 1 MW 3 GeV
protons from a rapid cycling synchrotron on a mercury target with 25 Hz repetition 
in the MLF.
The experiment uses a Gadolinium (Gd) loaded liquid scintillator
(Gd-LS) detector with 0.1 \% Gd concentration placed at
24 m from the target. In addition, di-isopropylnaphthalene
(DIN,C$_{16}$H$_{20}$) was dissolved into the Gd-LS by 8-10\% concentration in volume from 2021.

The JSNS$^2$ experiment aims to directly test the LSND
observation \cite{CITE:LSND} using improvements on the experimental technique.
Observing $\bar{\nu}_{\mu} \rightarrow \bar{\nu}_e$ oscillation using a
$\mu$DAR neutrino source via inverse beta decay (IBD) reaction,
$\bar{\nu}_e+p\rightarrow e^++n$, is the same experimental principle
used by LSND experiment \cite{CITE:LSND}.
On the other hand, there are several improvements offered by
the JSNS$^2$ experiment. The main improvements come from using a beam 
with a low duty factor and a Gd-loaded liquid scintillator, which reduces the accidental background significantly~\cite{CITE:JSNS2TDR}. 

There are also a number of correlated backgrounds to the IBD signal,
characterized by time-coincident prompt and delayed events.
The most concerning ones are cosmic-induced neutrons whose prompt
signal is made by a recoil proton and a delayed signal from the
neutron capture on Gd after thermalization. 
The first data of JSNS$^2$ taken in the commissioning phase in 2020 proves this facts~\cite{CITE:EPJC}. Therefore, approximately 99\% of this neutron background should be rejected by Pulse Shape Discrimination (PSD) using the waveform shape difference between the prompt signal of IBD and neutron events. The PSD technique will be applied to the full fiducial volume to reject the neutron background, thus a crucial improvements have been done from the previous report ~\cite{CITE:EPJC}. The JSNS$^2$ experiment PSD technique, based on a data driven likelihood method, will be described in this manuscript.


\section{\label{sec:detector} JSNS$^2$ detector}

The JSNS$^2$ detector is described elsewhere~\cite{CITE:JSNS2NIM}. However, in this section, we discuss the most relevant points regarding PSD. 

The JSNS$^2$ experiment features a cylindrical liquid scintillator detector with 4.6 m diameter and 3.5 m height placed at a distance of 24 m from the mercury target of the MLF. It consists of
17 tonnes of Gd-LS contained in an acrylic vessel, and 33
tonnes unloaded liquid scintillator (LS) in a layer between the
acrylic vessel and a stainless steel tank. The LS and the Gd-LS
consist of LAB (linear alkyl benzene) as the base solvent, 3
g/L PPO (2,5-diphenyloxazole) as the fluor, and 15 mg/L bis-MSB (1,4-bis(2-methylstyryl) benzene) as the wavelength
shifter. The DIN was dissolved into the Gd-LS in the
acrylic vessel which has dimensions of 3.2 m of diameter and
2.5 m of height in order to enhance the PSD capability.
DIN is commercially available and widely used as a base solvent
of organic liquid scintillator. Several neutrino experiments
using a liquid scintillator detector have adopted it and achieved
strong PSD capability~\cite{CITE:NEOS,CITE:JINST}.
Approximately 8\% by volume of DIN was dissolved into the
Gd-LS at the beginning of the first physics run from January
2021. Nitrogen purging was performed before data taking
and nitrogen gas is flowed into the gas phase of the detector
to avoid oxygen contamination from outside.
Starting from the physics run in 2022, the concentration of DIN was increased from 8\% to 10\%. 

The LS volume is separated into two independent layers by an optical separator that forms two subvolumes
in the one detector. The region inside the optical separator,
called the “inner detector”, consists of the entire volume of
the Gd-LS and $\sim$25 cm thick LS layer. Scintillation light
from the inner detector is observed by 96 Hamamatsu R7081
photomultiplier tubes (PMTs) each with a 10-inch diameter.
The outer layer, called the “veto layer”, is used to detect
cosmic-ray induced particles coming into the detector. A total
of 24 10-inch PMTs are set in the veto layer whose inner
surfaces are fully covered with reflection sheets in order to
improve the collection efficiency of the scintillation light.

PMT signal waveforms from both the inner detector and the
veto layer are digitized and recorded at a 500 MHz sampling
rate by 8-bit flash analog-to-digital converters (FADCs)~\cite{CITE:JSNS2DAQ}. A trigger, called the “self trigger”, uses an analog sum of the PMT signals from the inner detector. The trigger threshold is set such that the detection efficiency is approximately 100\% above 7 MeV. The FADCs record 496 ns wide waveforms during the data taking using the self trigger. We utilized the self trigger 
in order to obtain cosmogenic events, and when taking calibration
data using a 252-Californium ($^{252}$Cf) neutron source. The self-trigger is used for the PSD performance study described in this manuscript.

\section{\label{sec:ES} Dataset and event selection}

The analysis used 6 days of data which were taken with 8\% volume of DIN dissolved GdLS at the beginning of 2021. The relationship between DIN concentration and PSD capability is under study. Before going into the details of the PSD technique, the event selections of the cosmogenic Michel electrons (ME) 
and fast neutrons (FN) are described below. These control samples are used to make probability density 
functions (PDFs) of the likelihood and to evaluate the PSD capabilities which will be described later.
The selection criteria for those samples are shown in Table \ref{tab:EventSelection}. 
In addition to the energy and timing selections, 
a spatial correlation between prompt and delayed signal (ME: less than 130 cm, and FN: less than 60 cm) is adopted.


\begin{figure}[h]
\centering
\includegraphics[width=0.6\textwidth]{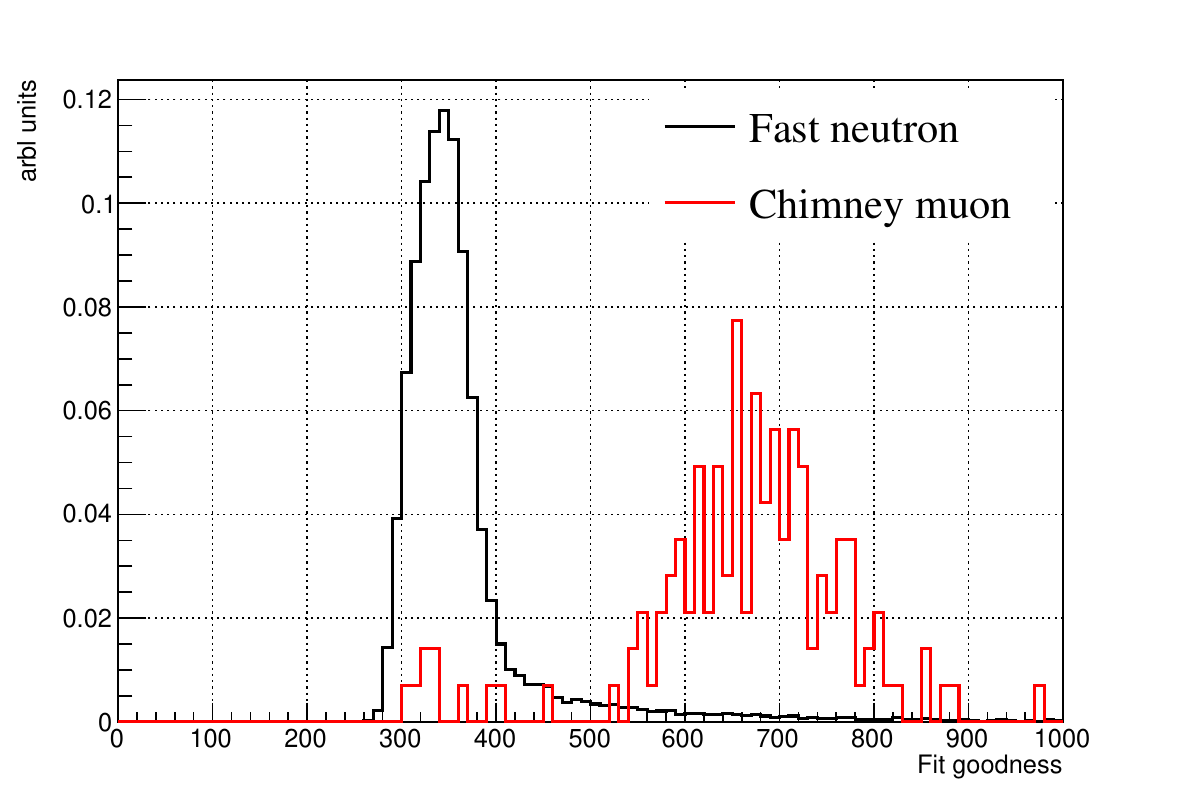} 
\caption{\label{fig:Recm} The fit goodness of fast neutrons and chimney muons by event reconstruction in JSNS$^2$.} 
\end{figure}

To reject cosmic muons for ME delayed and FN samples, we require that the top 12 and bottom 12 veto PMTs have less than 100 photoelectrons (p.e.). In JSNS$^{2}$, the energy and the vertex are reconstructed by JADE (JSNS$^{2}$ Analysis Development Environment), which is based on a maximum likelihood technique ~\cite{CITE:Johnathon, CITE:JSNS2reco}. A fit goodness event reconstruction by JADE is used to improve the purity of the control sample; events with a fit goodness variable $>$ 500 are rejected. When muons pass through the chimney which is located at the top center of the detector target, veto PMTs will not register hits from the muons. Further these muons do not initiate isotropic scintillation light inside the chimney, thus a higher goodness value is returned as shown in Fig. \ref{fig:Recm}. Hence, these muons are removed efficiently using the fit goodness variable.
The fiducial volume is defined with $R$ $< 140$ cm and $|z| <$ 100 cm region to avoid 
external backgrounds. Note that origin of the coordinate system is the center of the detector, 
and $R$ is defined as $R = \sqrt{x^2 + y^2}$.

\begin{table}
\centering

\begin{tabular}{c|c|c}\hline\hline
&ME&FN\\\hline\hline
Prompt energy&10 - 800 MeV&20 - 60 MeV\\
Dealyed energy&20 - 60 MeV&7 - 12 MeV\\
Time coincidence ($\Delta t$)&2 $< \Delta t < 10$$\mu$s&2 $< \Delta t < 100$$\mu$s\\
Spatial coincidence ($\Delta_{VTX}$)&$\Delta_{VTX}$ $<$ 130 cm&$\Delta_{VTX}$ $<$ 60 cm\\
Fit goodness&$<$500&$<$500\\
\hline\hline
\end{tabular}

\caption{\label{tab:EventSelection} The selection criteria for control samples.
		Cosmogenic Michel electrons (ME) delayed signal and neutron (FN) prompt signals are used as
        control samples.} 
\end{table}

With these selections, 387713 ME and 21623 FN events are available for the PSD evaluation and 243310 ME and 15160 FN events are used to make PDFs in this manuscript.
Note that the ME sample contains both electrons and positrons and the relevant energy region is identical to the IBD sample.

\section{\label{sec:principle} Principle of PSD}

\begin{figure}[h]
\centering
\includegraphics[width=0.6\textwidth]{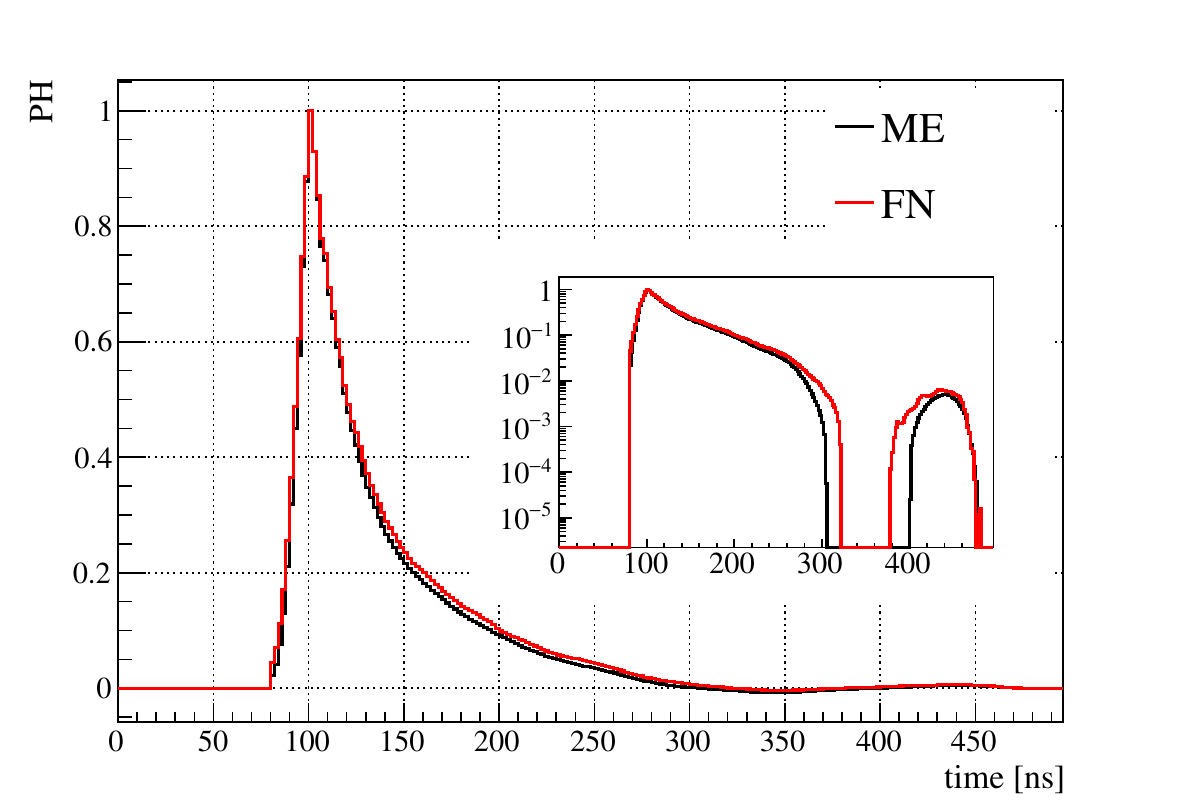} 
\caption{\label{fig:PSD} The typical averaged PMT waveforms of cosmogenic Michel electrons 
(black) and neutron (red) events.} 
\end{figure}

In this section, the principle of the PSD technique is described.
The scintillation light created from particle with large local energy loss ($dE/dx$) creates larger timing
tails than those from minimum ionizing particles.
Fig.~\ref{fig:PSD} shows the typical averaged PMT waveforms of cosmogenic Michel electrons 
(black) and neutron events (red), which are observed by the JSNS$^2$ detector. The selection criteria  
will be described later.
As shown in this Fig.~\ref{fig:PSD}, the waveforms created by the recoiled protons of neutrons
have larger tails than that of Michel electrons. PSD will use this difference. 

\subsection{\label{sec:pdf} Probability Density Function (PDF)}

\begin{figure}[h]
\centering
\includegraphics[width=0.9\textwidth]{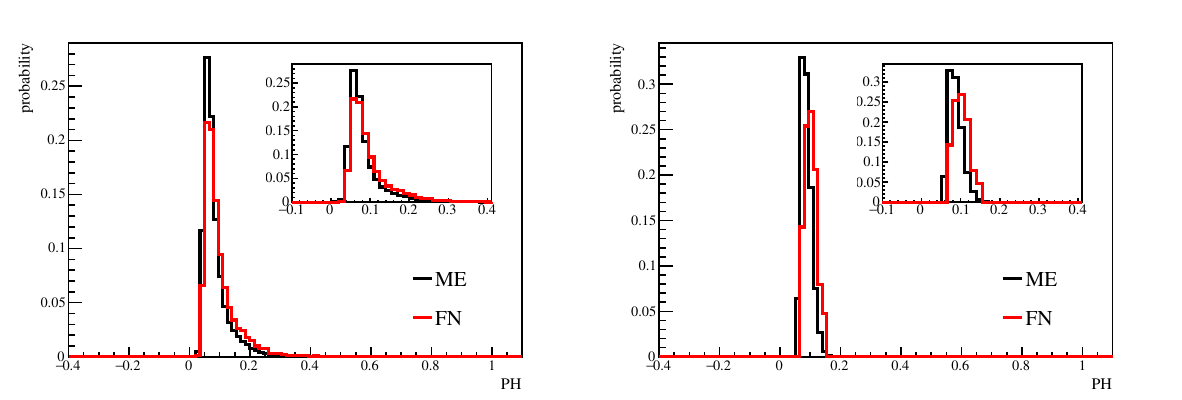} 
\caption{\label{fig:PDF} Typical PDFs of one PMT with around 200 ns of 
waveforms. Left: a low charge case. Right: a large charge case.} 
\end{figure}

For making the likelihoods for PSD, probability density functions (PDFs) utilizing 
the pulse height of each FADC bin are created. As shown in Fig.~\ref{fig:PSD}, waveforms are recorded every 2 ns, 
thus the peak normalized pulse height of each 2 ns slice is used to create the
PDFs. In the tails, the pulse heights made from ME are 
distinct from FN events; this difference is utilized for PSD. Each PMT has its own waveform characteristics depending on
the received charge but also the power to discriminate the ME and FN alone, 
therefore the charge dependent PDFs are made for each PMT separately. 
Note that the large PMT charge creates small fluctuations on the shape of the waveforms. 
Figure~\ref{fig:PDF} shows a typical PDF of one PMT using around 200 ns of the waveform.
Left shows the low charge (20 $<$ Q $<$ 40 p.e.) and right shows the 
high charge (200 $<$ Q $<$ 220 p.e.) cases, respectively. Only data are used to make the PDFs.

The waveforms before the real signal, i.e.: before around 80 ns in Fig.~\ref{fig:PSD},
is unused for the PSD algorithm. The low charge (Q $<$ 20 p.e.) or high charge (Q $>$ 500 p.e.) PMTs are also neglected in order to avoid unstable waveforms.

\subsection{\label{sec:evaluation} Evaluation}

The likelihood score is calculated with the following equation:
\begin{equation}
	\mathcal{L} =  \prod_{i=0}^{95} \prod_{j=40}^{247} \left[P_{ij}(PH)\right]~~,
\end{equation}
where $i$ is the PMT number of the JSNS$^2$ inner detector, which consists of 96 PMTs, 
$j$ is the index of the FADC time bin (40-247 bins in the self-trigger) such as 
shown in Fig.~\ref{fig:PSD}, and $PH$ is the peak normalized pulse height in $j$-th bin.
$P_{j}(PH)$ is calculated from the PDFs shown in Fig.~\ref{fig:PDF}. Note that PMTs, which show effects from the parent muon in their waveforms are rejected in the calculation.

There are two PDFs for the ME and FN, thus the likelihood ratio is calculated
with the PDFs as follows:
\begin{equation}
	\mathcal{L_{R}}= \frac{\mathcal{L^{ME}}}{\mathcal{L^{FN}}} =  \prod_{i=0}^{95} \prod_{j=40}^{247}
    \frac{\left[P^{ME}_{ij}(PH)\right]}{\left[P^{FN}_{ij}(PH)\right]} 
\label{eq:LR}
\end{equation}

For the convenience of the calculation, the log-likelihood is used. Equation \ref{eq:LR} 
is transformed to 
\begin{eqnarray*}
  	ln\mathcal{(L_{R})} &=& ln\mathcal{(L^{ME})} - ln\mathcal{(L^{FN})} \\
    &=& \sum_{i=0}^{95} \sum_{j=40}^{247} ( ln (P^{ME}_{ij}(PH) ) - ln (P^{FN}_{ij}(PH) ) ) 
\end{eqnarray*}

This log-likelihood ratio provides a positive score for events which are more electron like, 
while events with negative scores are more neutron like.

\section{\label{sec:performance} PSD performance}

\begin{figure}[h]
\centering
\includegraphics[width=0.65\textwidth]{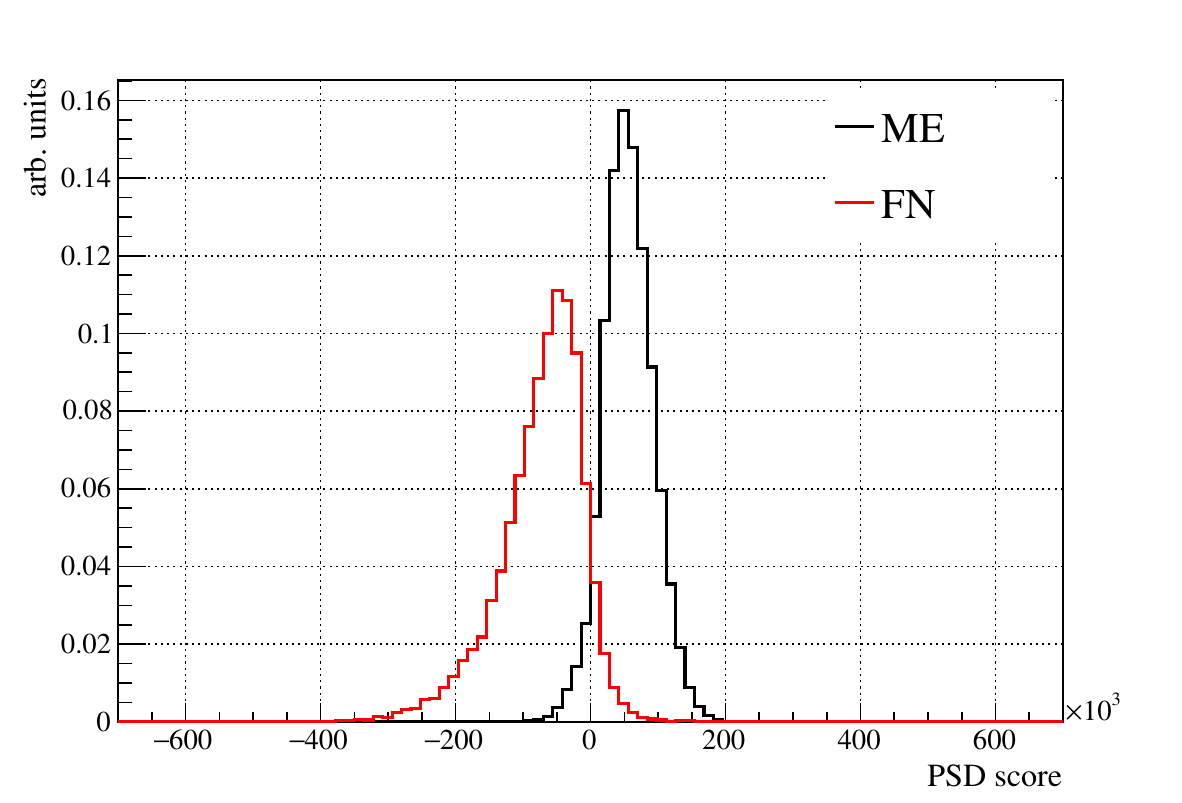} 
\caption{\label{fig:PSDscore} PSD score distributions on the log-likelihood ratio.} 
\end{figure}

In this section, the performance of the JSNS$^2$ PSD is described. To avoid bias, the control sample for making the PDFs and evaluating performance 
are separated and different dataset are used.
Fig.~\ref{fig:PSDscore} shows the PSD score distributions. As mentioned, events with positive score
are electron like events, while those with negative score are neutron like events. The black corresponds to ME and the red to the FN control samples using the event selection shown in Table~\ref{tab:EventSelection}. 

\begin{figure}[h]
\centering
\includegraphics[width=0.65\textwidth]{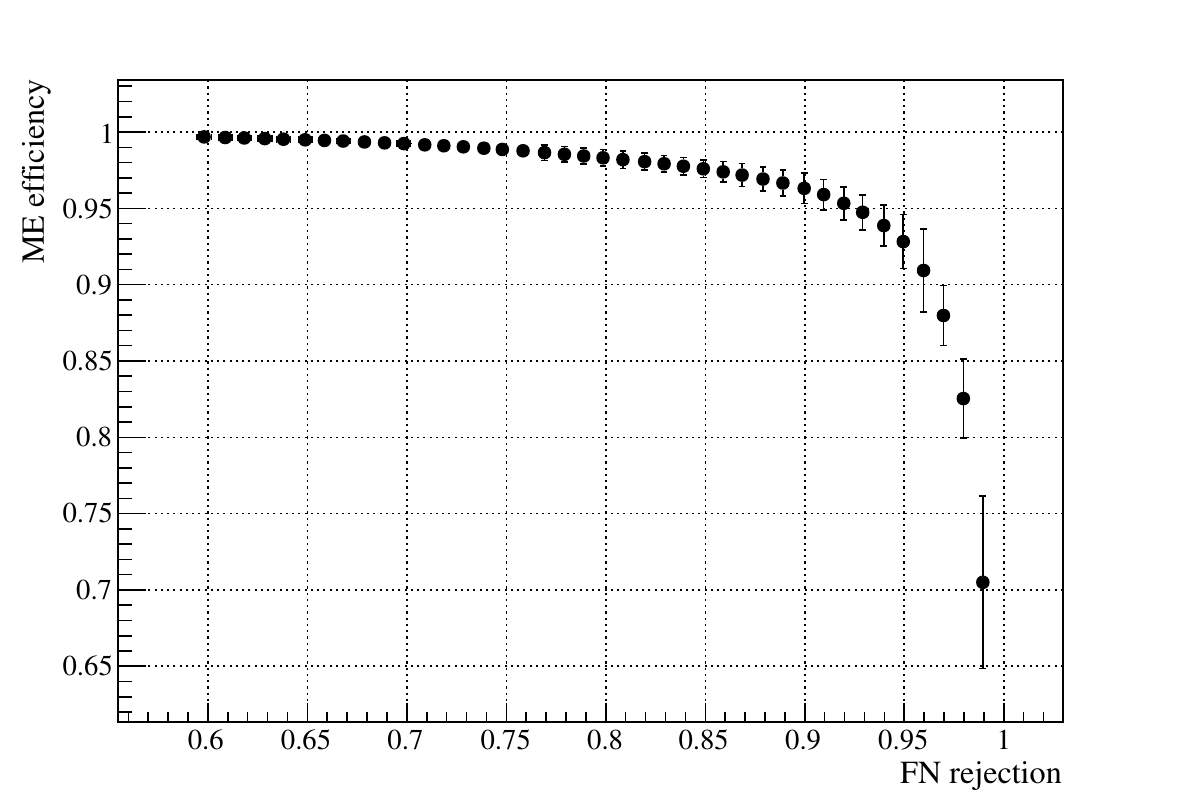} 
\caption{\label{fig:PSDperformance} The relationship between the FN rejection factor (horizontal) and ME efficiency (vertical).} 
\end{figure}

The relationship between the FN rejection factor and ME efficiency is seen in Fig.~\ref{fig:PSDperformance}. Requiring 94.95$\pm$0.15\% of FN rejection, the ME efficiency is about 92.82$\pm$1.77\%, with the uncertainties including systematics.

\section{\label{sec:conclusion} Conclusion}

JSNS$^2$ has developed a log-likelihood based PSD to differentiate neutron background from signal electron-like events. Due to the DIN dissolved liquid scintillator and sophisticated likelihood techniques, described in this paper, we achieved 94.95$\pm$0.15\% neutron event rejection while keeping the signal like electron efficiency of 92.82$\pm$1.77\%, noting that the rejection factor can be increased depending on the analyses. JSNS$^2$ will adopt this PSD technique
for in searching for short-baseline oscillations involving a sterile neutrino.

\section*{Acknowledgement}
\justifying
We deeply thank the J-PARC staff for their supports, 
especially for the MLF and accelerator groups to provide 
the good opportunities of this experiment. We acknowledge the support of the Ministry of Education, Culture, ports, Science, and Technology (MEXT) and the JSPS grants-in-aid: 16H06344, 16H03967, 24K17074, 23K13133 and 20H05624, Japan. This work is also supported by the National Research Foundation of Korea (NRF): 2016R1A5A1004684, 17K1A3A7A09015973, 017K1A3A7A09016426, 2019R1A2C3004955, 2016R1D1A3B02010606, 017R1A2B4011200, 2018R1D1A1B07050425, 2020K1A3A7A09080133, 020K1A3A7A09080114, 2020R1I1A3066835, 2021R1A2C1013661, 2022R1A5A1030700,
2021R1A6A1A03043957 and RS-2023-00212787. Our work has also been supported by a fund from the BK21 of the NRF. The University of Michigan gratefully acknowledges the support of the Heising-Simons Foundation. This work conducted at Brookhaven National Laboratory was supported by the U.S. Department of Energy under Contract DE-AC02-98CH10886. The work of the University of Sussex is supported by the Royal Society grant no. IESnR3n170385. We also thank the Daya Bay Collaboration for providing the Gd-LS, the RENO collaboration for providing the LS and PMTs, CIEMAT for providing the splitters, Drexel University for providing the FEE circuits and Tokyo Inst. Tech for providing FADC boards.

\nocite{*}





\providecommand{\noopsort}[1]{}\providecommand{\singleletter}[1]{#1}%

\end{document}